\documentclass[doublecol]{epl2} 

\usepackage{bm,graphicx,graphics,amsmath,amssymb,bm,epsfig,color,pifont}
\usepackage{euscript,tabularx,textcomp}
\usepackage[english]{babel}

\title{Magnetization relaxation in the single molecule magnet Ni$_4$\\
  under continuous microwave irradiation}
\shorttitle{Magnetization relaxation in the single molecule magnet Ni$_4$}

\author{G. de Loubens\inst{1} \and D. A.  Garanin\inst{2} \and C. C.
  Beedle\inst{3} \and D. N.  Hendrickson\inst{3} \and A. D.
  Kent\inst{1}} 
\shortauthor{G. de Loubens \etal}

\institute{
  \inst{1} Department of Physics, New York University, 4 Washington
  Place, New York, New York 10003, USA\\
  \inst{2} Department of Physics and Astronomy, City University of New
  York, 250 Bedford Park Boulevard West, Bronx, New York 10468-1589, USA\\
  \inst{3} Department of Chemistry and Biochemistry, University of
  California San Diego, La Jolla, California 92093, USA
}

\pacs{75.45.+j}{Macroscopic quantum phenomena in magnetic systems}
\pacs{75.50.Xx}{Molecular magnets}
\pacs{76.30.-v}{Electron paramagnetic resonance and relaxation}

\abstract{Spin relaxation between the two lowest-lying spin-states has
  been studied in the $S=4$ single molecule magnet Ni$_4$ under steady
  state conditions of low amplitude and continuous microwave
  irradiation. The relaxation rate was determined as a function of
  temperature at two frequencies, 10 and 27.8~GHz, by simultaneously
  measuring the magnetization and the absorbed microwave power. A
  strong temperature dependence is observed below 1.5~K, which is not
  consistent with a direct single-spin-phonon relaxation process. The
  data instead suggest that the spin relaxation is dominated by a
  phonon bottleneck at low temperatures and occurs by an Orbach
  mechanism involving excited spin-levels at higher temperatures.
  Experimental results are compared with detailed calculations of the
  relaxation rate using the universal density matrix equation.}
\begin{document}
\bibliographystyle{eplbib}

\maketitle


Quantum tunneling of magnetization (QTM) \cite{friedman96,thomas96}
and quantum phase interference \cite{wernsdorfer99,ramsey08} have been
intensively studied in single molecule magnets (SMMs). These materials
have also been suggested as candidates for qubits in quantum
processors \cite{leuenberger01} and for applications in molecular
spintronics \cite{bogani08}. Quantum manipulation of the spin in such
materials can be envisioned provided they have sufficiently long spin
relaxation ($T_1$) and dephasing times ($T_2$). A first step has been
the study of coherent QTM, in which the tunneling rates are faster
than the rate of decoherence \cite{hill03,barco04b}. These studies
provided a lower bound on $T_2$ of about $0.5$~ns. In SMM crystals
dipolar interactions may limit the dephasing time, making it necessary
to work with dilute SMM ensembles \cite{hallak07, loubens08,
  henderson} or even individual molecules. For instance, a recent
study of dilute doped antiferromagnetic wheels demonstrated a phase
relaxation time of several microseconds at 1.8~K \cite{ardavan07}.

The phase relaxation rate $\Gamma_2=1/T_2$ is generally much greater
than the energy relaxation rate $\Gamma$. For instance, an upper bound
on $T_2$ of 50~ns was found in dilute Ni$_4$ solutions at 130~GHz and
5.5~K, where energy relaxation, rather than dipolar or hyperfine
interactions, may be the limiting mechanism \cite{loubens08}.
Interestingly, the relaxation rate $\Gamma$ due to a direct
spin-phonon process of a SMM embedded in an elastic medium can be
derived without any unknown coupling constant \cite{chudnovsky04,
  chudnovsky05}. Moreover, collective relaxation effects are expected
in SMM single crystals, such as phonon superradiance
\cite{chudnovsky04a}.

Direct measurements of the magnetization under pulse microwave (MW)
irradiation enable study of spin-dynamics in SMM crystals
\cite{petukhov07,bahr08,bal06a,bal08}. In Fe$_8$, long pulses
($>10~\mu$s) of resonant MW radiation drive spins and phonons out of
equilibrium, and heating effects or phonon bottleneck (PB) preclude
the observation of fast dynamics \cite{bal06a, petukhov07}. However,
using very short high power MW pulses, refs.~\cite{bahr08} and
\cite{bal08} showed that these effects can be circumvented, enabling
the observation of spin-dynamics on microsecond time scales.
Ref.~\cite{bahr08} explains the observed temperature dependence of the
level lifetimes with direct spin-phonon coupling whereas
ref.~\cite{bal08} concludes that a PB develops and plays an essential
role in the magnetization dynamics.

Several experiments on SMM crystals in the quasi-static regime
\cite{chiorescu00, schenker05} were explained based on the \emph{ad
  hoc} model of PB \cite{faughnan61, scott62}. Only recently has this
effect been investigated on microscopic basis \cite{garanin07,
  garanin08}. When $N_S$, the number of spins in a crystal, becomes
comparable to $N_\mathrm{ph}$, the number of resonant phonon modes
(which can take energy out from the spin system), spin relaxation
leads to the so-called bottleneck plateau \cite{garanin07}. Only
because phonons themselves can relax to another reservoir is thermal
equilibrium eventually attained. Hence, the collective dynamics of
spins \emph{plus} resonant phonons depends on phonon relaxation, which
can be limited at low temperatures \cite{lounasmaa74}. In SMM crystals
the bottleneck parameter $B\equiv N_S/N_\mathrm{ph}$ is very large
($\gg 10$), even though the resonant lines are broadened
by inhomogeneities \cite{garanin08}. Hence, collective effects in
spin-phonon relaxation should be strong in SMMs.


This Letter presents a study of magnetization relaxation in the SMM
Ni$_4$ under steady state conditions of low amplitude and continuous
microwave irradiation. The general methods used can be
  applied to any spin system, including other molecular magnets. A
large transverse field is used to tune the energy splitting between
the two lowest spin-states \cite{barco04b}, while small deviations
from equilibrium magnetization induced by a resonant MW field are
monitored. Simultaneous measurements of the absorbed MW power and of
the magnetization enable measurement of energy relaxation rate as a
function of temperature. Experimental results are compared with
detailed calculations of the relaxation rate between the first excited
state and the ground state, using the density matrix equation (DME)
with the relaxation terms in the universal form \cite{chudnovsky04,
  chudnovsky05}.


In steady state, the energy level populations are constant. Under
continuous wave (cw) irradiation of frequency $f=\omega_0/(2\pi)$, the
absorbed power $P_\mathrm{abs}$ is proportional to the spin-photon
transition rate $\Lambda$ and to the population difference
$N_0=N_a-N_b$ between the states $|a\rangle$ and $|b\rangle$ involved
in the transition. The latter is reduced from its equilibrium value
$N_0^\mathrm{eq}$ by the resonant perturbation.  This reduction
depends on the ratio between the transition and relaxation rates. For
a two level system, the total relaxation rate is
$\Gamma=\Gamma_{ab}+\Gamma_{ba}$ and
$P_\mathrm{abs}=\hbar\omega_0\Lambda N_0=\hbar\omega_0\Lambda
N_0^\mathrm{eq}/(1+2\Lambda/\Gamma)$. $\Lambda$ can be eliminated
giving $\Gamma=2P_\mathrm{abs}/(\hbar\omega_0(N_0^\mathrm{eq}-N_0))$.
The populations can be obtained from magnetization measurement, $M=N_a
m_a+N_b m_b=M^\mathrm{eq}+\Delta M$. Thus the total relaxation rate
between $|a\rangle$ and $|b\rangle$ can be expressed in terms of the
measurable quantities $P_\mathrm{abs}$ and $\Delta M$ as follows:
\begin{equation}
  \Gamma=\frac{(m_b-m_a)}{\Delta M}\frac{P_\mathrm{abs}}{\hbar\omega_0}.
  \label{relaxP-DM}
\end{equation}
Eq.~(\ref{relaxP-DM}) is a simple result of the balance between
absorption and relaxation in steady state for a two level system. If
the system has more than two levels -- with levels $a$ and $b$ lowest
in energy -- one can still consider an effective two-level model in
which relaxation between $|a\rangle$ and $|b\rangle$ is assisted by
upper levels, provided that the temperature is sufficiently low so
that the populations of all other levels are small. The use of
eq.~(\ref{relaxP-DM}) is still justified because the upper levels do
not contribute to $P_\mathrm{abs}$ and $\Delta M$.


\begin{figure}
  \includegraphics[width=8.5cm]{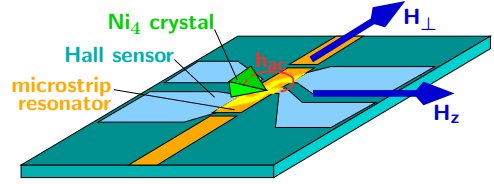}
  \caption{(Color online) Schematic of the integrated sensor used in
    this study.}
  \label{Fig1}
\end{figure}

Simultaneous measurements of $P_\mathrm{abs}$ and $\Delta M$ have been
previously used to measure relaxation times in paramagnets and
ferromagnets \cite{bloembergen54, loubens05}. Recently, quantitative
determination of the magnetization suppression under high power cw
irradiation was obtained in the SMM Fe$_8$ using a commercial SQUID
\cite{cage05}. Here, we use a novel integrated sensor that
incorporates a microstrip resonator with a micro-Hall magnetometer
\cite{loubens07b}, which enables low power studies. Briefly, the Hall
sensor is fabricated from a 2 dimensional electron gas in a
GaAs/AlGaAs heterostructure. A microstrip resonator is capacitively
coupled to two transmission lines and lithographically placed on the
Hall sensor, as shown in fig.~\ref{Fig1}. A thin dielectric separates
the resonator from the heterostructure material. The magnetometer,
biased with a dc current of $10~\mu$A, permits detection of changes in
the magnetization $\Delta M$ as small as $10^{-4}$ of the saturation
magnetization $M_s$ of our SMM crystals. The resonator has a
fundamental resonance frequency of 10~GHz, for which the ac field
$h_\mathrm{ac}$ is maximum at the sample location. This is also the
case for the third harmonic, at 27.8~GHz. The large filling factor of
the resonator enables the detection of absorbed power in the nW range
for $\approx 100^3\mu$m$^3$ samples.  $h_\mathrm{ac}$ was calibrated
at room temperature at each frequency through DPPH saturation. A
vector network analyzer is used as a MW source and for transmission
measurements. The integrated sensor is mounted in a $^3$He cryostat
with a base temperature of 0.37~K, in a superconducting vector magnet
that enables magnetic fields to be applied in arbitrary directions
with respect to the axes of the crystal.


[Ni(hmp)(dmb)Cl]$_4$ (Ni$_4$) is a particularly clean SMM with no
solvate molecules present in its crystal phase and only 1\% (natural
abundance) of nuclear spins on the transition metal sites
\cite{yang03}. This results in narrower EPR peaks than in many SMMs
\cite{edwards03}. The spin Hamiltonian of Ni$_4$ is to first
approximation:
\begin{equation}
  \hat{H}_S=-DS_z^2-BS_z^4+C(S_+^4+S_-^4)-\mu_B\bm{H}\cdot\hat{\bm g}\cdot\bm{S},
  \label{ham}
\end{equation}
where the first term is the uniaxial anisotropy, the second and third
terms are $4^{th}$-order anisotropy terms, and the last term is the
Zeeman energy. The $S=4$ ground state of the molecule at low
temperature is a consequence of ferromagnetic exchange interactions
between the four Ni$^{\mathrm{II}}$ ($S=1$) ions. The uniaxial
anisotropy leads to a large energy barrier $DS^2\approx12$~K to
magnetization reversal between states of projection $S_z=\pm4$ along
the easy axis of the molecule. The large $C$--term is at the origin of
the fast tunneling ($\approx 10$~MHz) observed at zero field
\cite{kirman05} and a recent analysis shows that the $4^{th}$-order
terms in eq.~(\ref{ham}) result from a finite ratio of the (second
order) single ion anisotropies to exchange constant \cite{wilson06}.


\begin{figure}
  \includegraphics[width=8.5cm]{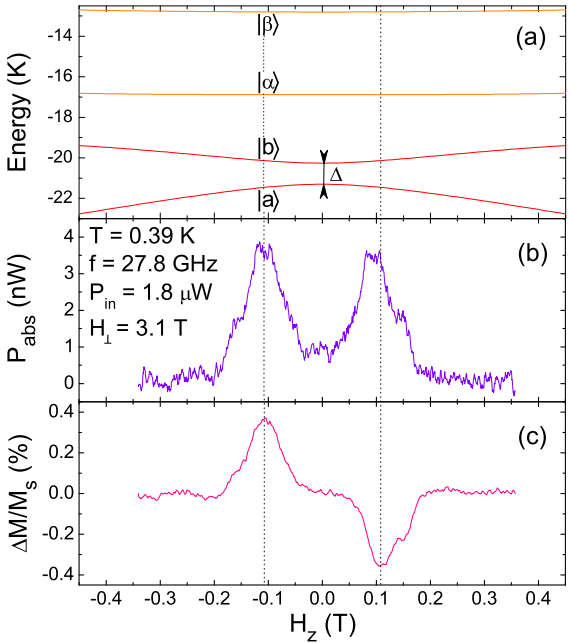}
  \caption{(Color online) (a) Energy levels of Ni$_4$ vs. longitudinal
    field $H_z$ in the presence of a transverse field $H_\perp=3.1$~T. (b)
    Absorbed power and (c) magnetization changes vs. $H_z$ measured
    with an the integrated sensor as the Ni$_4$ crystal is irradiated
    at 27.8 ~GHz. The dashed vertical lines show the resonant
    transitions between the two lowest levels $|a\rangle$ and
    $|b\rangle$ corresponding to this frequency.}
  \label{Fig2}
\end{figure}

A pyramidal Ni$_4$ crystal containing $N_S=(7\pm0.5)\times10^{14}$
molecules is affixed to the integrated sensor with vacuum grease and
oriented so that $h_\mathrm{ac}$ is within
  $20$\textdegree~of $z$, the axis of the pyramid. A constant
transverse field $H_\perp$ is applied\footnote{At an angle
    $\phi\simeq11$\textdegree~with the hard axis, estimated from the
    crystal geometry \cite{kirman05}.} perpendicularly to $z$, while
the longitudinal field $H_z$ is swept
\cite{barco04b,loubens07b}. In this configuration, an
avoided crossing between the two lowest energy levels $|a\rangle$ and
$|b\rangle$ occurs, with a tunnel splitting $\Delta$
(fig.~\ref{Fig2}(a)). The energy levels and eigenstates are calculated
by direct diagonalization of the Hamiltonian (eq.~(\ref{ham})). The
character of the states $|a\rangle$ and $|b\rangle$ changes with
$H_z$. At $H_z=0$, they have zero magnetization along $z$. To have a
measurable magnetization we work at finite $H_z$, where $m_a$ and
$m_b$ are also finite and almost opposite. In the presence of cw MW
irradiation, power absorption can be detected when the splitting
between $|a\rangle$ and $|b\rangle$ matches the photon energy
$\hbar\omega_0$ (fig.~\ref{Fig2}(b)). Simultaneous magnetization
changes $\Delta M$ are observed
(fig.~\ref{Fig2}(c))\footnote{Ni$_4$ molecules in a
    different molecular environment are responsible for the peak
    shoulders at $\approx \pm 0.15$~T \cite{edwards03}.}. By
repeating this experiment at different transverse fields with the MW
frequency fixed, it is possible to determine the zero field splitting
parameters \cite{barco04b, loubens07b}, which are found to be in good
agreement with those deduced from high frequency EPR experiments
\cite{edwards03, kirman05}.  Using the two frequencies available for
our microstrip resonator, we obtain $D=0.75$~K, $B=7\times10^{-3}$~K,
$C=2.9\times10^{-4}$~K, $g_x=g_y=2.23$ and $g_z=2.3$.

\begin{figure}
  \includegraphics[width=8.5cm]{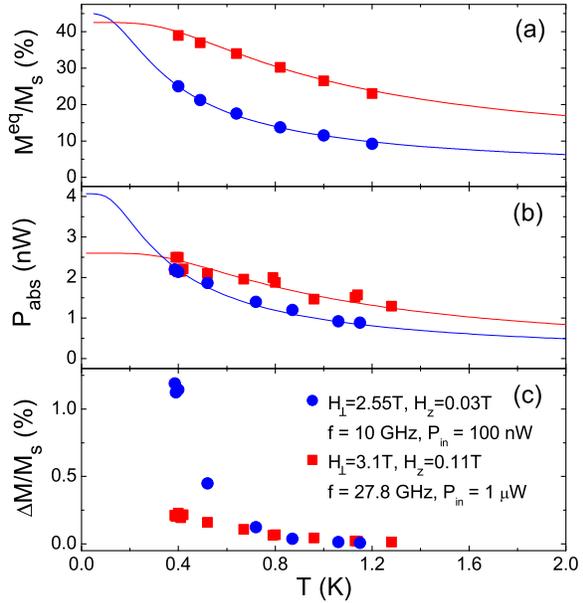}
  \caption{(Color online) (a) Equilibrium magnetization, (b) absorbed
    power and (c) magnetization changes vs. temperature, measured for
    two values of the applied field, corresponding to transition
    frequencies of 10 and 27.8~GHz. The continuous lines are
    calculations (see text).}
  \label{Fig3}
\end{figure}

We now concentrate on the behavior of $P_\mathrm{abs}$ and $\Delta M$
vs. temperature for two values of the applied field (fig.~\ref{Fig3}).
$(H_\perp,H_z)_1=(2.55, 0.03)$~T corresponds to a transition frequency of
10~GHz and $(H_\perp,H_z)_2=(3.1, 0.11)$~T to 27.8 GHz. In both cases,
only the two lowest-lying states are significantly populated within
the temperature range investigated. At 1.5~K, less than 3\% of the
molecules occupy other states.  The first such state lies several
Kelvins above $b$ (see fig.~\ref{Fig2}(a)), and thus the effective
two-state model discussed above is valid. The temperature dependence
of the equilibrium magnetization in the absence of MW,
$M^\mathrm{eq}$, is plotted in fig.~\ref{Fig3}(a). It is well
reproduced by the two-state model (continuous lines):
\begin{equation}
  \frac{M^\mathrm{eq}}{M_s}=\frac{m_a+m_b \exp{(-\hbar\omega_0/(k_B T))}}{S(1+\exp{(-\hbar\omega_0/(k_B T))})},
  \label{Meq}
\end{equation}
where $m_a$, $m_b$ and $\omega_0$ are calculated from diagonalization
of $\hat{H}_S$. It is found that $m_a/S=0.45$ and $m_b/S=-0.432$ for
$(H_\perp,H_z)_1$, and that $m_a/S=0.425$ and $m_b/S=-0.329$ for
$(H_\perp,H_z)_2$.

When irradiating the sample with MW, the power was deliberatly kept
small in order to investigate small deviations from equilibrium,
typically less than 1\%. The lowest achievable sample temperature
under these conditions is 0.385 K, only 15 mK above the cryostat base
temperature. The measured temperature dependence of $P_\mathrm{abs}$
is plotted in fig.~\ref{Fig3}(b) together with the formula:
\begin{equation}
  P_\mathrm{abs}=\hbar\omega_0 \left(\frac{g\mu_B}{\hbar}h_\mathrm{ac}\right)^2 \frac{|\langle a|S_z|b\rangle|^2}{2\Gamma_2^*} N_S \tanh{\left(\frac{\hbar\omega_0}{2k_B T}\right)}.
  \label{Pabs}
\end{equation}
$\Gamma_2^*$ is an upper bound on the dephasing rate, and
can be extracted from the line width, found to be about 4~GHz
\cite{loubens07b}. The only fitting parameter for the continuous lines
displayed in fig.~\ref{Fig3}(b) is $h_\mathrm{ac}$, which is found to
be 1.9~mOe at 10~GHz and 1.3~mOe at 27.8~GHz. These values are 30 to
50\% less than expected from calibration. The difference in the
temperature dependences of $M^\mathrm{eq}$ and $P_\mathrm{abs}$ under
the different applied fields is simply due to the different level
splittings, $\hbar \omega_0$, in the Boltzmann factors of
eqs.~(\ref{Meq}) and (\ref{Pabs}). Finally $\Delta M$, the
magnetization change induced by the MW absorption, is plotted in
fig.~\ref{Fig3}(c) for the two applied fields. Its temperature
dependence is much stronger than that of $M^\mathrm{eq}$ and
$P_\mathrm{abs}$. $\Delta M/M_s$ decreases by almost two orders of
magnitude at 10~GHz as $T$ increases from 0.385 to 1.2~K, and by
almost a factor 20 at 27.8~GHz.

\begin{figure}
  \includegraphics[width=8.5cm]{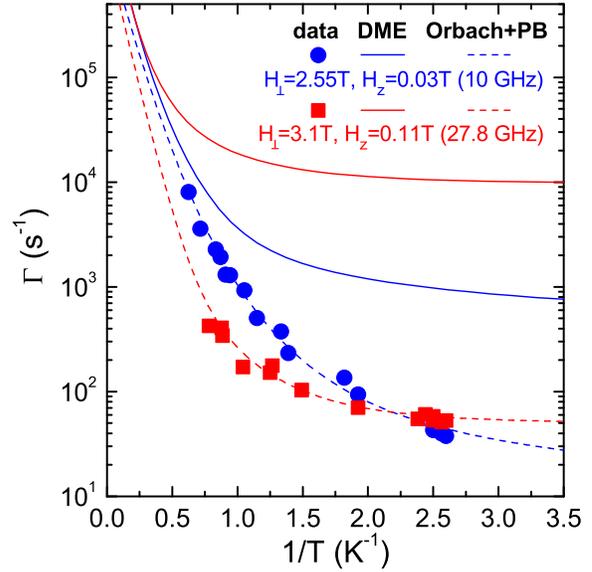}
  \caption{(Color online) Dependence of the energy relaxation rate at
    10 and 27.8~GHz on inverse temperature deduced from the
    simultaneous measurements of fig.~\ref{Fig3} and from
    eq.~(\ref{relaxP-DM}). The continuous lines are calculations using
    the universal DME and the dashed lines are fits of the data
    including the Orbach mechanism and PB (see text).}
  \label{Fig4}
\end{figure}

With the data presented in fig.~\ref{Fig3} and eq.~(\ref{relaxP-DM})
we have extracted\footnote{In order to extract $\Gamma$ up to
  $T=1.6$~K, the 10~GHz data of fig.~\ref{Fig4} also include
  measurements at a MW power of $P_\mathrm{in}=1~\mu$W.} the energy
relaxation rate $\Gamma$. It is displayed in a semilog plot vs.  the
inverse temperature $1/T$ in fig.~\ref{Fig4}. The general trend of the
data at 10 and 27.8~GHz is a moderate increase of $\Gamma$ with
temperature at low temperatures ($T< 0.8$~K) followed by an
exponential growth at higher temperatures.

A direct spin-phonon relaxation process from $|b\rangle$ to
$|a\rangle$ is not consistent with the data, because it would lead to
a more gradual temperature dependence of the relaxation rate. In the
case where the transverse field is mainly responsible for the energy
splitting between states, it reads \cite{chudnovsky05},
\begin{equation}
  \Gamma_{\mathrm{sp},T}=\frac{1}{2}QS^2\frac{\Delta^2\omega_0(g\mu_B H_\perp)^2}{12\pi
    E_t^4}\mathrm{coth}\left(\frac{\hbar\omega_0}{2k_B T}\right),
  \label{spinphonon}
\end{equation}
where $Q\cong1$ when $SH_z<H_\perp$ and $E_t=(\rho
v_t^5\hbar^3)^{1/4}$ is the characteristic energy of the spin-phonon
interaction. $\rho$ is the density of Ni$_4$, and $v_t$ the speed of
the transverse phonons. The temperature dependence of
$\Gamma_{\mathrm{sp},T}$ is much weaker than that of the measured rate
$\Gamma$, even at low temperature. The zero temperature spin-phonon
relaxation rate $\Gamma_\mathrm{sp}$ given by the prefactor of
eq.~(\ref{spinphonon}) is very sensitive to the precise value of the
transverse speed of sound, since $v_t$ appears to the fifth power.
The latter was extracted from the heat capacity $C(T)$ of Ni$_4$
\cite{hendrickson05} using the extended Debye model \cite{garanin08a}.
The soft phonon mode, $v_t=940$~m/s, is nearly transverse and the most
important for the spin-phonon relaxation, whereas the two stiff modes
with $v=2250$~m/s can be neglected. The prefactor
  $\frac{1}{2}$ in eq.~(\ref{spinphonon}) accounts for the fact that
  only one phonon mode contributes to the spin-phonon process. Using
the density of Ni$_4$, $\rho=1.4$~g/cm$^{3}$ \cite{yang06}, we obtain
$E_t=75.7$~K. Using this value in eq.~(\ref{spinphonon}),
the expected zero temperature relaxation rate from the direct process
is found to be 40 and 200 times larger than the observed rate at
10~GHz and 27.8~GHz, respectively. Such a large disagreement can not
be assigned to an uncertainty of $v_t$, which would have to differ by
a factor 2 to 3 from the value we have taken.  Also processes other
than rotation of magnetic molecules without distortion, unaccounted
for in the universal form of the spin-phonon coupling, cannot explain
the discrepancy since they can only increase the relaxation
rate.

This suggests that a PB is necessary to explain the data. In fact, an
excellent agreement with the data at each frequency is obtained with
the form of relaxation rate that combines PB and Orbach mechanism (see
dashed lines in fig.~\ref{Fig4}):
\begin{equation}
  \Gamma_T=\frac{\Gamma_\mathrm{ph}}{B_{\omega_{0}}}\mathrm{coth}^2\left(\frac{\hbar\omega_0}{2k_B T}\right)+\sum_i{\Gamma_{0i}\exp{\left(\frac{-E_i}{k_BT}\right)}}.
  \label{PB-Orbach}
\end{equation}
This temperature dependence of the relaxation rate has also been
observed in dilute paramagnetic salts \cite{ruby62}.


The temperature dependence of the PB mechanism in the presence of
phonon relaxation $\Gamma_\mathrm{ph}$ can be obtained from eqs.~(60)
of ref.~\cite{garanin07} by including a large inhomogeneous
broadening, as done in ref.~\cite{garanin08}. In that case, the
bottleneck parameter has the form:
\begin{equation}
  B_{\omega_{0}}\equiv\frac{N_{S}}{N}\frac{\rho_{S}(\omega_{0})}{\rho_{\mathrm{ph}}(\omega_{0})}=\frac{N_S}{N\sqrt{2\pi}}\frac{2\pi^2\tilde{\Omega}_D^3}{\omega_0^2\delta\omega_0}.
  \label{Bparam}
\end{equation}
Here, the spin density of states $\rho_{S}(\omega_{0})$ is assumed to
have a Gaussian line shape of width $\delta\omega_0$, $N$ is the
number of unit-cells and $v_0$ their volume.
$\tilde{\Omega}_D=v_t/v_0^{1/3}$ is proportional to the Debye
frequency. Integrating out $r_\bold{k}$ and then over the narrow
natural line shape in eqs.~(60) of ref.~\cite{garanin07} with the
added inhomogeneous broadening leads to equations:
\begin{eqnarray}
  \dot{p}_{\omega_{0}} &=&-\Gamma_\mathrm{sp} \left[ p_{\omega_{0}}+\left( 2p_{\omega_{0}}-1\right) n_{\omega_{0}}\right]   \nonumber \\
  \dot{n}_{\omega_{0}} &=&\Gamma_{\mathrm{ph}}\left( n_{\omega_{0}}^{\mathrm{eq}}-n_{\omega_{0}}\right) -B_{\omega_{0}}\dot{p}_{\omega_{0}},
  \label{PBEqsInhomo}
\end{eqnarray}
that are similar to those of refs ~\cite{faughnan61,scott62}.
$p_{\omega_0}$ is the population of the spin's excited state and
$n_{\omega_0}$ the phonon population. From linearization of
eqs.~(\ref{PBEqsInhomo}) for $B_{\omega_{0}}\gg1$, it can be found
that the lowest rate governing the relaxation is:
\begin{equation}
  \Lambda_{-}\cong \frac{\Gamma_{\mathrm{ph}}\Gamma_{\mathrm{sp},T}}{B_{\omega_{0},T}\Gamma_{\mathrm{sp},T}+\Gamma_{\mathrm{ph}}},
  \label{Lambdamsmall}
\end{equation}
where $B_{\omega_{0},T}=B_{\omega_{0}}\tanh^2(\hbar\omega_0/(2k_B
T))$. In the strong bottleneck limit, $\Gamma_{\mathrm{ph}}\ll
B_{\omega_{0},T}\Gamma_{\mathrm{sp},T}$, the so-called bottleneck rate
$\Lambda_{-}\cong\Gamma_{\mathrm{b}}=\Gamma_{\mathrm{ph}}/B_{\omega_{0},T}$
dominates, as in our experiments at low temperature, \emph{i.e.}, the
first term of eq.~(\ref{PB-Orbach}). In Ni$_4$, the unit-cell contains
four molecules and its volume is $v_0=5777.1$~\AA$^3$ \cite{yang06}.
Hence, $N_S/N=4$ and $\tilde{\Omega}_D=8.3\times10^{11}$~s$^{-1}$.
Using $\delta\omega_0=25\times10^9$~s$^{-1}$,
$B_{\omega_{0}}\approx180000$ at 10~GHz and
$B_{\omega_{0}}\approx24000$ at 27.8~GHz, values which also depend
strongly on $v_t$. From the prefactors of the PB mechanism used to fit
the data (see dashed lines on fig.~\ref{Fig4}), we find that in our
experiments, the phonon relaxation time
$T_\mathrm{ph}=1/\Gamma_{\mathrm{ph}}\approx0.4~\mu$s at 10~GHz and
$\approx0.8~\mu$s at 27.8~GHz. This corresponds to a phonon mean free
path of about 500~$\mu$m, for a crystal whose largest dimension is
180~$\mu$m. These values are comparable to published ones
\cite{scott62}, and reveal that the phonons undergo several
reflections on the crystal walls before escaping.


To better understand the relaxation including the Orbach mechanism
that contributes to eq.~(\ref{PB-Orbach}), we have performed
calculations of the relaxation rate between the two lowest-lying
states in the whole temperature range, using the density matrix
equation (DME) with the relaxation terms in the universal form
\cite{chudnovsky04, chudnovsky05}. In this approach, the complete set
of energy levels and the effect of temperature are taken into account.
The explicit form of the crystal fields enters through the
eigenfunctions and eigenvalues of the Hamiltonian (eq.~(\ref{ham}))
that are universal building blocks of the relaxation term of the DME.
This makes inclusion of various anisotropies easy since it does not
change the structure of the latter. The continuous lines in
fig.~\ref{Fig4} are the results of DME calculations for the applied
fields $(H_\perp,H_z)_1$ and $(H_\perp,H_z)_2$ used in the
experiments. The Hamiltonian parameters and characteristic energy of
the spin-phonon coupling needed for the calculation are the same as
given above. The low temperature part of the DME calculations
corresponds to the direct spin-phonon process, which is given by
eq.~(\ref{spinphonon}). As previously explained, this process is
screened out by PB. The calculated relaxation rate follows the direct
process until the temperature reaches about 0.8~K.  Above this
temperature, an exponential growth of the rate is observed.  This
behavior would not be expected for a pure two-level system. But within
the giant spin approximation, the $S=4$ Ni$_4$ SMM has seven other
levels lying above the two lowest ones, $|a\rangle$ and $|b\rangle$.
Two of them, $|\alpha\rangle$ and $|\beta\rangle$, have been
represented in fig.~\ref{Fig2}(a). Although they are weakly populated
within the temperature range investigated, their presence can greatly
enhance the relaxation rate. Instead of the direct relaxation
$|b\rangle \rightarrow |a\rangle$, the system can be thermally excited
from $|b\rangle$ to some high level, say $|i\rangle$, and then fall
down to $|a\rangle$.  This is the Orbach mechanism, yielding a
characteristic Arrhenius temperature dependence of the rate,
\emph{i.e.}, the second term of eq.~(\ref{PB-Orbach}). Its activation
energy $E_i$ equals the separation between levels $b$ and $i$, whereas
its strength $\Gamma_{0i}$ depends on the coupling between $|i\rangle$
and $|a\rangle$ through transverse phonons.

\begin{table}
  \begin{center}
    \begin{tabular}{|c|c|c|c|}
      \hline
      level $i$ & 3 & 4 & 5\\
      \hline
      10~GHz, $E_i$~(K) & 3.27 & 6.45 & 10.61\\
      $\Gamma_{0i}^\mathrm{DME}$~(s$^{-1}$) & $1.7\times10^4$ & $5\times10^5$ & $1.5\times10^6$\\
      $\Gamma_{0i}^\mathrm{fit}$~(s$^{-1}$) & $10^4$ & $2.5\times10^5$ & $1.5\times10^6$\\
      \hline\hline
      27.8~GHz, $E_i$~(K) & 3.25 & 7.34 & 12.28\\
      $\Gamma_{0i}^\mathrm{DME}$~(s$^{-1}$) & $3.5\times10^4$ & $7\times10^5$ & $1.5\times10^6$\\
      $\Gamma_{0i}^\mathrm{fit}$~(s$^{-1}$) & $2\times10^3$ & $5\times10^4$ & $1.5\times10^6$\\
      \hline
    \end{tabular}
  \end{center}
  \caption{Levels involved in the Orbach mechanism for the two studied
    frequencies. The calculated strengths using DME as well as the
    fitted values to the data are summarized.}
  \label{tab1}
\end{table}

From the same statistical argument mentioned in the introduction, the
Orbach mechanism can also be bottlenecked. But the emitted phonons
have a much higher frequency $E_i/h$ than those from the direct
process, leading to a much lower bottleneck parameter (\emph{i.e.},
there is a much higher density of phonon modes available for spin
relaxation).  Assuming the same inhomogeneous broadening
$\delta\omega_0$, $B_{\omega_0}\approx800$ for $E_i=7$~K, which is
close to the energy separation between $|b\rangle$ and the fourth
energy level for the two different applied fields. Assuming that
$\Gamma_\mathrm{ph}$ is independent of temperature, only processes
with a rate larger than about $5\times 10^3$~s$^{-1}$ will be strongly
bottlenecked (eq.~(\ref{Lambdamsmall})). This can be seen from the
data, which behave differently at 10 and at 27.8~GHz in the high
temperature region of fig.~(\ref{Fig4}). Whereas the measured rate at
27.8~GHz is still two orders of magnitude lower than the calculated
one, the data at 10~GHz tend to meet the DME calculation. This is due
to the fact that as the Orbach mechanism starts, the rate computed at
10~GHz using the universal DME is still low
($\Gamma\approx10^3-10^4$~s$^{-1}$), and hence weakly bottlenecked. On
the contrary, the relaxation at 27.8~GHz is still strongly
bottlenecked in that temperature range because the calculated rate is
faster than $10^4$~s$^{-1}$. Table~\ref{tab1} summarizes the three
Orbach channels relevant for the energy relaxation in the temperature
range studied. It gives their activation energies, their calculated
strengths using DME, and the values fitted to the data using
  eq.~(\ref{PB-Orbach}). The fitted values are smaller than the
calculated ones because PB is not completely suppressed. As explained,
this effect is stronger at 27.8~GHz than at 10~GHz.


It is interesting to estimate the phonon relaxation rate necessary to
suppress the PB from eq.~(\ref{Lambdamsmall}). If the phonon
relaxation is very fast, $B_{\omega_{0},T}\Gamma_{\mathrm{sp},T}\ll
\Gamma_{\mathrm{ph}}$, $\Lambda_{-}\cong \Gamma_{\mathrm{sp},T}$, and
there is no bottleneck. One would need to have $T_\mathrm{ph}\ll
10$~ns at 10~GHz and $T_\mathrm{ph}\ll4$~ns at 27.8~GHz to meet this
requirement at low temperature. This would only be possible for a
submicron-size crystal.

In conclusion, the temperature dependence of the energy relaxation
rate between the two lowest-lying spin-states of Ni$_4$ has been
measured at two different frequencies. Using calculations of the rate
based on the universal DME and taking into account PB that screens out
the direct spin-phonon process at low temperature, the two sets of
data are well accounted for in the whole temperature range
investigated. In particular, the Orbach mechanism dominates at high
temperatures, and is weakly bottlenecked at 10~GHz. Finally, we point
out that the fact that the universal DME calculations and the PB work
so well to explain the data is surprising, as coherence is expected to
play an important role in collective spin-lattice relaxation in SMM
crystals \cite{chudnovsky04a,calero07a}.

\acknowledgments This work was suppported by NSF Grants No.
DMR-0506946 and DMR-0703639. We aknowledge E. \textsc{del Barco},
S. \textsc{Hill} and E. M. \textsc{Chudnovsky} for useful
discussions.


\begin{thebibliography}{10}
\expandafter\ifx\csname url\endcsname\relax\def\url#1{\texttt{#1}}\fi

\bibitem{friedman96}
\Name{Friedman J.~R., Sarachik M.~P., Tejada J. \and Ziolo R.} \REVIEW{Phys.
  {R}ev. {L}ett.}{76}{1996}{3830}.

\bibitem{thomas96}
\Name{Thomas L., Lionti F., Ballou R., Gatteschi D., Sessoli R. \and Barbara
  B.} \REVIEW{Nature}{383}{1996}{145}.

\bibitem{wernsdorfer99}
\Name{Wernsdorfer W. \and Sessoli R.} \REVIEW{Science}{284}{1999}{133}.

\bibitem{ramsey08}
\Name{Ramsey C.~M., del Barco E., Hill S., Shah S.~J., Beedle C.~C. \and
  Hendrickson D.~N.} \REVIEW{Nature Physics}{4}{2008}{277}.

\bibitem{leuenberger01}
\Name{Leuenberger M.~N. \and Loss D.} \REVIEW{Nature}{410}{2001}{789}.

\bibitem{bogani08}
\Name{Bogani L. \and Wernsdorfer W.} \REVIEW{Nature Materials}{7}{2008}{179}.

\bibitem{hill03}
\Name{Hill S., Edwards R.~S., Aliaga-Alcalde N. \and Christou G.}
  \REVIEW{Science}{302}{2003}{1015}.

\bibitem{barco04b}
\Name{del Barco E., Kent A.~D., Yang E.~C. \and Hendrickson D.~N.}
  \REVIEW{Phys. {R}ev. {L}ett.}{93}{2004}{157202}.

\bibitem{hallak07}
\Name{Hallak F.~E., van Slageren J., Gomez-Segura J., Ruiz-Molina D. \and
  Dressel M.} \REVIEW{Phys. {R}ev. {B}}{75}{2007}{104403}.

\bibitem{loubens08}
\Name{de~Loubens G., Kent A.~D., Krymov V., Gerfen G.~J., Beedle C.~C. \and
  Hendrickson D.~N.} \REVIEW{J. Appl. Phys.}{103}{2008}{07B910}.

\bibitem{henderson}
\Name{Henderson J.~J., Ramsey C.~M., Datta S., del Barco E., Hill S., Stamatatos T.~C. \and Christou
  G.} in preparation.

\bibitem{ardavan07}
\Name{Ardavan A., Rival O., Morton J. J.~L., Blundell S.~J., Tyryshkin A.~M.,
  Timco G.~A. \and Winpenny R. E.~P.} \REVIEW{Phys. {R}ev. {L}ett.}{98}{2007}{057201}.

\bibitem{chudnovsky04}
\Name{Chudnovsky E.~M.} \REVIEW{Phys. Rev. Lett.}{92}{2004}{120405}.

\bibitem{chudnovsky05}
\Name{Chudnovsky E.~M., Garanin D.~A. \and Schilling R.} \REVIEW{Phys. {R}ev.
  {B}}{72}{2005}{094426}.

\bibitem{chudnovsky04a}
\Name{Chudnovsky E.~M. \and Garanin D.~A.} \REVIEW{Phys. Rev. Lett.}{93}{2004}{257205}.

\bibitem{petukhov07}
\Name{Petukhov K., Bahr S., Wernsdorfer W., Barra A.-L. \and Mosser V.}
  \REVIEW{Phys. {R}ev. {B}}{75}{2007}{064408}.

\bibitem{bahr08}
\Name{Bahr S., Petukhov K., Mosser V. \and Wernsdorfer W.} \REVIEW{Phys. Rev. B}{77}{2008}{064404}.

\bibitem{bal06a}
\Name{Bal M., Friedman J.~R., Rumberger E.~M., Shah S., Hendrickson D.~N.,
  Avraham N., Myasoedov Y., Shtrikman H. \and Zeldov E.} \REVIEW{J. Appl. Phys.}{99}{2006}{08D103}.

\bibitem{bal08}
\Name{Bal M., Friedman J.~R., Chen W., Tuominen M.~T., Beedle C.~C., Rumberger
  E.~M. \and Hendrickson D.~N.} \REVIEW{Europhys. Lett.}{82}{2008}{17005}.

\bibitem{chiorescu00}
\Name{Chiorescu I., Wernsdorfer W., Muller A., Bogge H. \and Barbara B.}
  \REVIEW{Phys. {R}ev. {L}ett.}{84}{2000}{3454}.

\bibitem{schenker05}
\Name{Schenker R., Leuenberger M.~N., Chaboussant G., Loss D. \and Gudel H.~U.}
  \REVIEW{Phys. {R}ev. {B}}{72}{2005}{184403}.

\bibitem{faughnan61}
\Name{Faughnan B.~W. \and Strandberg M. W.~P.} \REVIEW{J. {P}hys. {C}hem.
  {S}olids}{19}{1961}{155}.

\bibitem{scott62}
\Name{Scott P.~L. \and Jeffries C.~D.} \REVIEW{Phys. {R}ev.}{127}{1962}{32}.

\bibitem{garanin07}
\Name{Garanin D.~A.} \REVIEW{Phys. {R}ev. {B}}{75}{2007}{094409}.

\bibitem{garanin08}
\Name{Garanin D.~A.} \REVIEW{Phys. Rev. B}{77}{2008}{024429}.

\bibitem{lounasmaa74}
\Name{Lounasmaa O.~V.} \Book{Experimental principles and methods below 1 K}
  (Academic Press, London \& New York) 1974.

\bibitem{bloembergen54}
\Name{Bloembergen N. \and Wang S.} \REVIEW{Phys. Rev.}{93}{1954}{72}.

\bibitem{loubens05}
\Name{de~Loubens G., Naletov V.~V. \and Klein O.} \REVIEW{Phys. Rev. B}{71}{2005}{180411(R)}.

\bibitem{cage05}
\Name{Cage B., Russek S.~E., Zipse D., North J.~M. \and Dalal N.~S.}
  \REVIEW{Appl. {P}hys. {L}ett.}{87}{2005}{082501}.

\bibitem{loubens07b}
\Name{de~Loubens G., Chaves-O'Flynn G.~D., Kent A.~D., Ramsey C., del Barco E.,
  Beedle C.~C. \and Hendrickson D.~N.} \REVIEW{J. Appl. Phys.}{101}{2007}{09E104}.

\bibitem{yang03}
\Name{Yang E.-C., Wernsdorfer W., Hill S., Edwards R.~S., Nakano M., Maccagnano
  S., Zakharov L.~N., Rheingold A.~L., Christou G. \and Hendrickson D.~N.}
  \REVIEW{Polyhedron}{22}{2003}{1727}.

\bibitem{edwards03}
\Name{Edwards R.~S., Maccagnano S., Yang E.-C., Hill S., Wernsdorfer W.,
  Hendrickson D.~N. \and Christou G.} \REVIEW{Journ. {A}ppl. {P}hys.}{93}{2003}{7807}.

\bibitem{kirman05}
\Name{Kirman C., Lawrence J., Hill S., Yang E.-C. \and Hendrickson D.~N.}
  \REVIEW{J. Appl. Phys.}{97}{2005}{10M501}.

\bibitem{wilson06}
\Name{Wilson A., Lawrence J., Yang E.-C., Nakano M., Hendrickson D.~N. \and
  Hill S.} \REVIEW{Phys. {R}ev. {B}}{74}{2006}{140403}.

\bibitem{hendrickson05}
\Name{Hendrickson D.~N. \etal} \REVIEW{Polyhedron}{24}{2005}{2280}.

\bibitem{garanin08a}
\Name{Garanin D.~A.} \REVIEW{arXiv:0804.1066v3}{}{2008}{}.

\bibitem{yang06}
\Name{Yang E.-C.\etal} \REVIEW{Inorg. {C}hem.}{45}{2006}{529}.

\bibitem{ruby62}
\Name{Ruby R.~H., Benoit H. \and Jeffries C.~D.} \REVIEW{Phys. {R}ev.}{127}{1962}{51}.

\bibitem{calero07a}
\Name{Calero C., Chudnovsky E.~M. \and Garanin D.~A.} \REVIEW{Phys. Rev. B}{76}{2007}{094419}.

\end{thebibliography}

\end{document}